\documentstyle[prl,aps]{revtex}

\begin{document}

\preprint{Preprint}
\draft{}
\twocolumn[\hsize\textwidth\columnwidth\hsize 
           \csname @twocolumnfalse\endcsname   

\title{A reconstruction from small-angle neutron scattering measurements of 
the real space magnetic field distribution in the mixed 
state of Sr$_{2}$RuO$_{4}$.}

\author{P.~G.~Kealey$^{1,*}$,
T.~M.~Riseman$^{1}$,
E.~M.~Forgan$^{1}$,
L.~M.~Galvin$^{1}$,
A.~P.~Mackenzie$^{1}$,
S.~L.~Lee$^{2}$, 
D.~M$^c$K.~Paul$^{3}$,
R.~Cubitt$^{4}$,
D.~F.~Agterberg$^{5}$,
R.~Heeb$^{6}$,
Z.~Q.~Mao$^{7}$,
Y.~Maeno$^{7}$.
}

\address{$^{1}$School of Physics and Astronomy, University of 
Birmingham,
Birmingham B15 2TT, UK. \\
$^{2}$School of Physics and Astronomy, University of St.~Andrews,
St.~Andrews, Fife KY16 9SS, UK.\\ 
$^{3}$Department of Physics, University of Warwick, Coventry CV4 7AL,
UK.\\
$^{4}$Institut Laue-Langevin, 38042 Grenoble Cedex, France. \\
$^{5}$National High Magnetic Field Laboratory, Florida State University, Tallahassee, FA 32306, USA.\\
$^{6}$Theoretische Physik, ETH Honggerberg, CH-8093 Zurich, Switzerland. \\
$^{7}$Department of Physics, Kyoto University, Kyoto 606-8052, Japan.\\}
\date{\today}
\maketitle
\widetext

\begin{abstract}
We have measured the diffracted neutron scattering intensities from the square 
magnetic flux lattice in the perovskite superconductor Sr$_2$RuO$_4$, which is thought
to exhibit p-wave pairing with a two-component order parameter. The relative intensities
of different flux lattice 
Bragg reflections over a wide range of field and temperature have been shown to 
be inconsistent with a single component Ginzburg-Landau theory but
qualitatively agree with a two component p-wave Ginzburg-Landau theory.

\end{abstract}
%\pacs{DRAFT VERSION: NOT FOR DISTRIBUTION}
]
\narrowtext 
The discovery of superconductivity at temperatures near 1K in strontium 
ruthenate \cite{srofind} has excited great interest because it is a superconducting layered
perovskite which does not contain copper. However it shows great differences 
from the High-T$_c$ cuprates: it is a stoichiometric {\em undoped} 
compound with a long 
mean free path, in which the electrons form a Fermi liquid with a 
well-established quasi-two-dimensional Fermi surface \cite{fermisurface}. Furthermore, it was suggested \cite{pwave} that the 
strongly interacting electrons pair in a triplet p-wave state (rather than the 
singlet, mainly d-wave state which is believed to occur in hole-doped cuprates).
Clear evidence of non s-wave pairing in this compound has been provided by the 
observation \cite{unconv} that {\em non}magnetic impurities strongly suppress T${_c}$, which 
extrapolates to $\approx 1.5$\,K in the clean limit. Strong support for
triplet (p-wave) 
pairing is given by the results of Ishida {\em et al.} who have measured the Knight 
shift with a field parallel to the RuO$_2$ planes \cite{knight}; the spin susceptibility 
measured by the Knight shift is not suppressed below T$_c$,unlike a singlet 
superconductor.  Also, ${\mu}$SR measurements in the Meissner state in zero
field \cite{timereverse} have revealed spontaneous 
fields, which can be generated by domain boundaries, 
surfaces and impurities in a superconductor which breaks time-reversal symmetry 
\cite{otherreverse}. Such states can arise most naturally with p-wave pairing, but also are 
possible with d-wave singlet pairing.

Agterberg \cite{agterberg1,agterberg2} argued that if the pairing was time-reversal symmetry breaking 
p-wave, then in tetragonal symmetry the {\bf d}-vector \cite{extrayoshi} has the symmetry
$\hat{\mathbf{z}}\cdot\exp (\pm\mathrm{i}\varphi)$ (${\varphi}$ is the azimuthal
angle about the tetragonal c axis), and a two-component Ginzburg Landau
(TCGL) 
theory would be expected to describe the superconductor. In zero field, this 
gives two degenerate states which are related by time reversal; with a field 
applied in the c-direction perpendicular to the planes, one is dominant, but 
the other is also present \cite{heeb}. Under these conditions, a small amount of 
anisotropy in the Fermi surface would lead to a square flux lattice instead 
of a triangular one, with the orientation of the square flux line lattice (FLL) relative to the 
crystal axes determined by the orientation of the fourfold anisotropy of the 
paired electrons. The FLL structure has been observed in 
this material \cite{sans} and is observed to be square over a wide range of field 
and temperature. The nearest-neighbour directions in the square FLL are at 
45${^\circ}$ to the Ru-O-Ru directions in the crystal lattice \cite{sanscorrection}.

These results are consistent with the pairing wavefunction described above. 
However, a square FLL is also seen in borocarbide superconductors, which 
are definitely non-p-wave \cite{don1,don2,wilde}.  Also, one can measure spontaneous fields in a 
superconductor by $\mu$SR due to other causes or from other states than that 
proposed, and application of a strong field in the basal plane to observe
the 
Knight shift might alter the pairing state. Hence, it is important to obtain 
further evidence as to what kind of superconductivity occurs in strontium 
ruthenate. Here we present a detailed study of the scattered neutron 
intensities from the FLL.  We show that they are not consistent with a single
component Ginzburg-Landau model.  Also we demonstrate how the local
$B({\mathbf{r}})$ may 
be reconstructed from our data and show that the FLL structure is quite 
different from the Abrikosov one.

We shall present measurements of intensities of higher-order Bragg reflections 
from the FLL so we consider how they are related to the FLL structure. The 
formula \cite{christen} for the integrated intensity $I_{hk}$ of a $(h,k)$ diffracted peak of 
wavevector ${\mathbf{q}}_{hk}$ gives: 

\begin{equation} \label{eq:intensities}
I_{hk}\propto \frac{F_{hk}^2}{q_{hk}},      
\end{equation}								
where $F_{hk}$ is a spatial Fourier component of the local field $B({\mathbf{r}})$ in the mixed 
state:

\begin{equation} \label{eq:fourier}
B({\mathbf{r}})=\sum_{h,k}F_{hk}\exp({\mathrm{i}}{\mathbf{q}}_{hk}\cdot{\mathbf{r}}).
\end{equation}
In the Abrikosov solution of the Ginzburg-Landau (GL) equations (as applied 
to a square lattice)~\cite{abrikosov}, the $F_{hk}$ are given by:

\begin{equation} \label{eq:abba}
F_{hk}\propto-(-1)^{(h^2+k^2+hk)}\cdot
\exp\bigg(-\frac{\pi}{2}(h^2+k^2)\bigg);
\end{equation}
this rapidly falls off with $q$ (see Table \ref{tab:tb1}).

The Abrikosov solution is only valid near $B_{c2}$. In high-$\kappa$ superconductors, with 
the field not close to $B_{c2}$, the London expression \cite{london} is appropriate instead. 
This gives $F_{hk}\propto 1/(1+q_{hk}^2\lambda^2)$. Note that unlike the Abrikosov solution,
all the $F_{hk}$ are positive. Table \ref{tab:tb1} shows that the Fourier
components fall off much less rapidly with
$q$. However, strontium ruthenate has a value of the Ginzburg Landau parameter 
$\kappa = \lambda/\xi \sim 2.0$ for the field along the c axis, which means 
that the London approach is not realistic except at 
{\em very} low inductions. Therefore, to see what conventional GL theory predicts for 
this material at lower fields, one must use the Brandt numerical solution of the GL equations 
\cite{brandtcal}. Typical results are given in Table \ref{tab:tb1}.

Next, we consider the Agterberg TCGL solution \cite{agterberg1,agterberg2}, which is equivalent 
to the Abrikosov one, except that there are two complex order parameters instead of 
only one. In the mixed state with $B$ parallel to c both components are 
automatically present because of mixed gradient terms in the free energy 
functional \cite{agterberg1,agterberg2}. Typical values from this theory for $F_{hk}$,
relative to $F_{10}$ and  the resulting SANS intensities are given in Table
\ref{tab:tb1}. 
It may be seen that the two-component theory gives intensities that fall off {\em much less rapidly}
with $q$ than those given by the one-component Abrikosov solution. 

Under the conditions of our experiments, where the field is not close to $B_{c2}$,
it may be argued that the Abrikosov approximation used by Agterberg is not 
appropriate. However, recently Heeb and Agterberg have solved numerically the 
GL equations at all fields for the TCGL case 
\cite{heeb2}. We also give in Table \ref{tab:tb1} a list of Fourier components from these 
calculations, using values of parameters that appear to describe 
our results quite well.

\begin{figure}
  \input{epsf}
  \epsfysize 5.5cm
  \centerline{\epsfbox{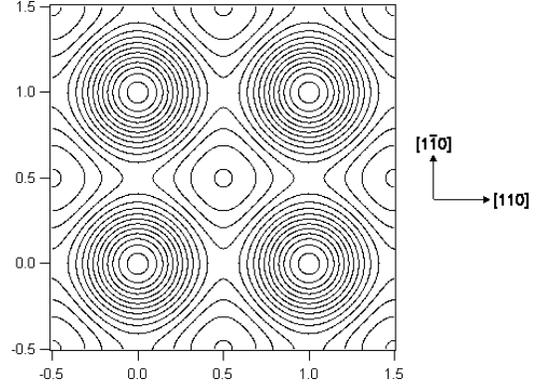}}
  \caption[~]{Contour plot of the magnetic field in a square flux line lattice
  as given by Brandt's numerical solution of the Ginzburg-Landau equations for the particular case of
  $\kappa =2.0,B=20mT$ and  $B_{c2}=58mT$. Contour lines are equally spaced.
  This result is very similar to that given by the Abrikosov solution valid near
  $H_{c2}$
  \cite{abrikosov}.} \label{fig:brandt}
\end{figure}

\begin{figure}
  \input{epsf}
  \epsfxsize 5.5cm
  \centerline{\epsfbox{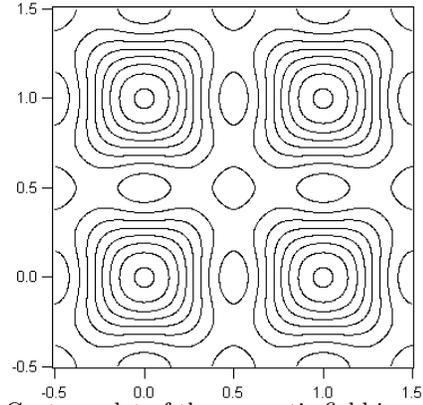}}
  \caption[~]{Contour plot of the magnetic field in 
a square flux line lattice as given by Agterberg's solution of his TCGL
equations, valid near $B_{c2}$, in the case of a cylindrical Fermi
surface ($\nu =0.0$).} \label{fig:agzero}
\end{figure}

\begin{figure}
  \input{epsf}
  \epsfxsize 5.5cm
  \centerline{\epsfbox{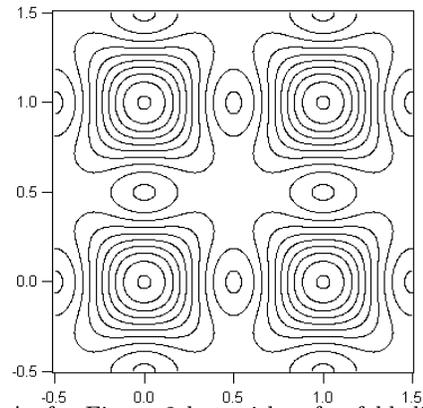}}
  \caption[~]{As for Figure \ref{fig:agzero} but with a fourfold
  distortion to the Fermi surface ($\nu =0.2$).}\label{fig:agp2}
\end{figure}

\begin{figure}
  \input{epsf}
  \epsfysize 5.5cm
  \centerline{\epsfbox{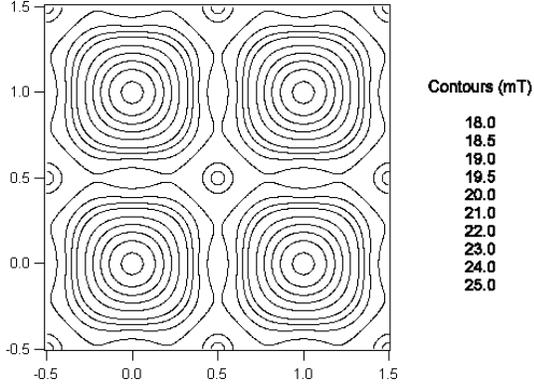}}
  \caption[~]{Heeb and Agterberg's numerical solution to the
  TCGL equations \cite{heeb2}, with the parameters $\kappa =1.6, \nu
=0.05$
chosen to give a good fit to our data, with an applied field of $B=20$mT
and $B_{c2}$(100mK)=58mT.}
\label{fig:heeb2} 
\end{figure}

The corresponding vortex structures in real space are shown in Figures
\ref{fig:brandt}-\ref{fig:heeb2}. 
Note that there is a minimum field point in the two-component theory (for the
conditions of our experiment) which lies {\em between} the positions of the flux line 
cores, not in the centre of the square. We give results of this theory for two values of the parameter $\nu$
($-1< \nu < 1$) \cite{agterberg2} 
which describes the degree of fourfold anisotropy of superconducting 
electrons ($\nu=0$ corresponds to a cylindrical Fermi surface).  We note that
the results do not change greatly with $\nu$. Hence the qualitative difference 
between Figures \ref{fig:brandt} and \ref{fig:agp2} is due to the difference
between TCGL and GL 
theories rather than effects of fourfold crystal anisotropy.  It may be 
that $\nu$ is quite small since $\vert\nu\vert >0.0114$ is sufficient to
stabilise a square FLL and align it to the crystal lattice with an orientation 
determined by the {\em sign} of $\nu$ \cite{agterberg2}.

\begin{figure}
  \input{epsf}
  \epsfysize 5.5cm
  \centerline{\epsfbox{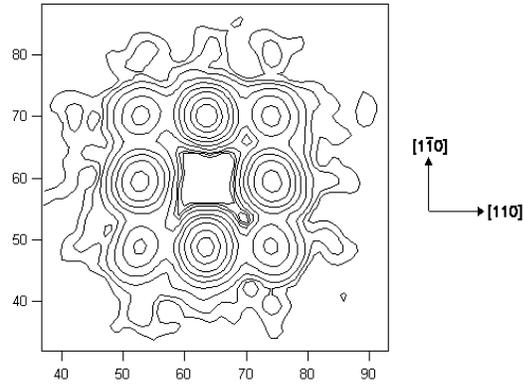}}
  \caption[~]{Contour plot of FLL diffraction pattern.  
 A field of 20mT was applied parallel to $c$ above T$_c$; the weak
diffracted beams due to the flux lattice were extracted from background
scattering by subtracting data taken above T$_c$. The axes are pixel numbers 
and the central region of the detector has been masked.  The data shown is a sum
of 5 patterns obtained by rocking the FLL up $\pm$0.3$^\circ$ about the [110]
axis.} \label{fig:fll} 
\end{figure}

We now turn to measurements of the FLL structure. Single crystal Sr$_2$RuO$_4$ 
was prepared by the floating zone technique with excess RuO$_2$ as a flux 
\cite{maeno96}. Six plates of total mass 556 mg were cleaved from the as-grown 
crystal and annealed for 72 hours in air at 1420${^\circ}$C to remove defects and
increase T$_c$, which was 1.39K with a width (10-90\%) of $\approx$50 mK.
With the 
field applied parallel to the c-axis at 100 mK, the value of $B_{c2}$ was 
58mT. For the small angle neutron scattering (SANS) measurements, the
samples
were mounted with conducting silver paint as an aligned mosaic with their c-axes perpendicular to a 
copper plate, which was mounted on the mixing chamber of a dilution 
refrigerator. This was placed between the poles of an electromagnet, 
which had holes parallel to the field for transmission of neutrons. The
magnetic field was parallel to the c-axes of the crystals within 2${^\circ}$, and
the FLL was observed using long-wavelength neutrons on instrument D22 at 
the Institut Laue Langevin. Typical wavelengths employed were 14.6\,\AA, 
with a wavelength spread (FWHM) of 12\%; the neutron beam was incident 
nearly parallel to the applied field, and the transmitted neutrons were 
registered at a $128\times 128$  pixel multidetector (pixel size $7.5\times7.5$\,mm$^2$) 
placed 17.71\,m beyond the sample. Typical results are shown in Fig
\ref{fig:fll}. In 
addition to the strongest $\{10\}$ reflections, the $\{11\}$ reflections
are strong, and higher orders are present. The intensity of the strongest 
diffraction spot is $<$10$^{-3}$ of the incident beam intensity, so these
higher 
order reflections are not due to multiple scattering. Their intensities 
are recorded in Table \ref{tab:tb1}: it will be noted that they are much larger than 
those given by the Abrikosov structure.

To reconstruct the $B({\mathbf{r}})$ of the FLL corresponding to these results, we 
require the sign of $F_{hk}$ relative to $F_{10}$ (the FLL is centrosymmetric, 
so all the $F_{hk}$ are real). The most important component after $F_{10}$ is $F_{11}$. 
If it has the same sign as $F_{10}$, then the $\{11\}$ components add in
phase 
at the flux line cores to give a field peak that is sharper than the 
field minimum. Measurements of the field distribution in strontium 
ruthenate by $\mu$SR \cite{musrfind,musrgeneral} show that this is the case. 
This sign for $F_{11}$ 
is not surprising, since all models in Table \ref{tab:tb1} give it as positive. 
For the small contributions of $F_{20}$ and $F_{21}$, we may assume the same signs 
as given by the Agterberg and Abrikosov solutions: taking the London 
sign makes a large difference to $B({\mathbf{r}})$, and also can be ruled out by 
$\mu$SR results . The reconstruction of $B({\mathbf{r}})$ is shown in Fig. 4. Note that 
it is completely different from the Abrikosov or Brandt solutions to 
the GL equations, and in good qualitative agreement with the TCGL 
predictions.  

The results we have given so far correspond to low temperature and a 
particular magnetic field value. In Table \ref{tab:tb2} we present the 
values of the form factors $F_{10}$ and $F_{11}$ for a range of 
fields at 100mK. Also, in Fig 3 we plot versus temperature 
the ratio of the Fourier components for the strongest two reflections 
$F_{11}/F_{10}$ at 10, 20 and 30mT.  Remarkably, this ratio varies little with 
field and temperature and does not tend to the Abrikosov value as T$\rightarrow$T$_c$.  
Non-local effects \cite{wilde} and deviations from GL theory in ultra pure
superconductors \cite{otherreason} should both die away at high temperatures. 
Therefore, these effects are not expected to be the cause of the flux line
shapes we report, although they may affect the details of $B$(r) at low
temperatures.

In conclusion, the strength of the higher order reflections from the FLL in 
strontium ruthenate and their temperature dependence certainly show 
that a standard one component Ginzburg Landau model is insufficient 
to explain the observed diffraction pattern. However, our results are in good
qualitative but not perfect agreement with a two component Ginzburg Landau
theory. Unconventional flux line 
shapes in this material are strong evidence for unconventional 
superconductivity in Sr$_2$RuO$_4$.

We thank J-L Ragazzoni of the ILL for setting up the dilution 
refrigerator, E.H. Brandt for a copy of his code to solve the GL 
equations and G.M. Luke for communicating his results prior to 
publication. This work was supported by the U.K. E.P.S.R.C., 
CREST of Japan Science and Technology Corporation, and the neutron 
scattering was carried out at the Institut Laue-Langevin, Grenoble.

\bibliographystyle{prsty}

\begin{thebibliography}{99}
\bibitem{srofind} Y.~Maeno {\em et al.}, Nature {\bf 372}, 532 (1994).
\bibitem{fermisurface}A.~P.~Mackenzie {\em et al.}, Phys. Rev. Lett., {\bf 76}, 3786 
(1996), C.Bergemann {\em et al.}, cond-mat/9909027
\bibitem{pwave} T.~M.~Rice and M.~Sigrist, J.Phys: Condens. Matter {\bf 7}, L643
(1995).
\bibitem{unconv} A.~P.~Mackenzie {\em et al.}, Phys. Rev. Lett., {\bf 80}, 161
and 3890 (1998).
\bibitem{knight} K.~Ishida {\em et al.}, Nature {\bf 396}, 658 (1998).
\bibitem{timereverse} G.~M.~Luke {\em et al.}, Nature {\bf 394}, 558 (1998).
\bibitem{otherreverse} M.~Sigrist and K.~ Ueda, Rev. Mod. Phys. {\bf 63}, 239
(1991).
\bibitem{agterberg1} D.~F.~Agterberg, Phys. Rev. Lett. {\bf 80}, 5184 (1998).
\bibitem{agterberg2} D.~F.~Agterberg, Phys. Rev. B. {\bf 58}, 14484 (1998).
\bibitem{extrayoshi} Y.~Maeno {\em et al.}, Journal of Superconductivity, {\bf 12}, 535
(1999).
\bibitem{heeb} R.~Heeb and D.~F.~Agterberg, Phys. Rev. B {\bf 59}, 7076 (1999)
\bibitem{sans}T.~M.~Riseman {\em et al.}, Nature {\bf 396}, 242-5 (1998).
\bibitem{sanscorrection}E.~M.~Forgan and D.~M$^c$K.~Paul, Correction to Nature,
to be published (2000).
\bibitem{don1} D.~M$^c$K.~Paul {\em et al.}. Phys. Rev. Lett. {\bf 80} 1517 (1998).
\bibitem{don2} M.~R.~Eskildsen {\em et al.}, Nature {\bf 393}, 242 (1998) 
\bibitem{wilde} Y.~De.~Wilde {\em et al.}, Phys. Rev. Lett. {\bf 78}, 4273, (1997).
\bibitem{christen} D.~K.~Christen {\em et al.}, Phys. Rev. B {\bf 15}, 4506-9
(1977).
\bibitem{abrikosov}  A.~A.~Abrikosov, 1957, Sov. Phys. JETP {\bf 5} 1174 (1957),
E.~H.~Brandt, Phys. Stat. Sol. B {\bf 64} 257, 467 {\bf 65}
469 (1974).
\bibitem{london} M.~Tinkham, 1975, Introduction to Superconductivity, Malabar, Florida, USA:McGraw-Hill 
\bibitem{brandtcal} E.~H.~Brandt, Phys. Rev. Lett., {\bf 78}, 2208 (1997).
\bibitem{heeb2} R.~Heeb and D.~F.~Agterberg, {\em to be published}
\bibitem{maeno96} Z.~Q.~Mao {\em et al.}, to be published in Mat. Res. Bull
(2000).
%Y.~Maeno {\em et al.}, J. Low temp Phys. {\bf 105}, 1577-1588(1996).
\bibitem{musrfind} C.M. Aegerter {\em et al.},  J.Phys: Cond. Mat., {\bf 10},
7445-51 (1998).
\bibitem{musrgeneral} G.~M.~Luke {\em et al.}, {\em to be published} (2000)
\bibitem{otherreason} J.~M.~Delrieu, J. Low Temp Phys., {\bf 6}, 197-219 (1972).
%\bibitem{heebprivate} R.~Heeb {\em Private communication}

\end{thebibliography}

$^{*}$\small{Email address: P.G.Kealey@bham.ac.uk}

\begin{figure}
  \input{epsf} \epsfysize 6cm 
  \centerline{\epsfbox{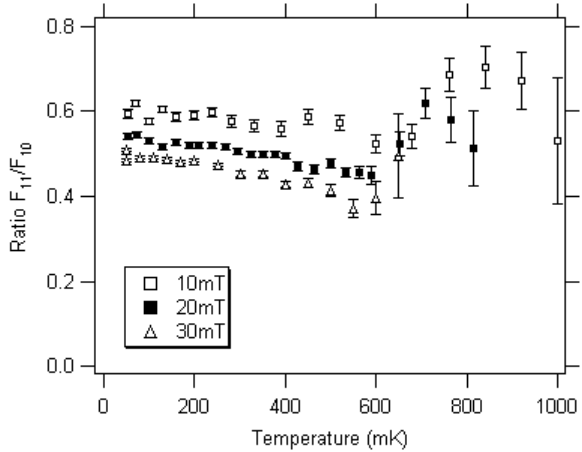}}
  \caption[~]{Temperature and field dependence of the ratio $F_{11}/F_{10}$.
The \{1,1\} intensity used is the direct average of all four spots, and the \{1,0\}
intensity is a weighted average of the top and side spots which allows for
the different rocking-curve width in the vertical and
horizontal directions \cite{sans} .} \label{fig:tdep}
\end{figure}
\begin{figure}
  \input{epsf}
  \epsfysize 5.5cm
  \centerline{\epsfbox{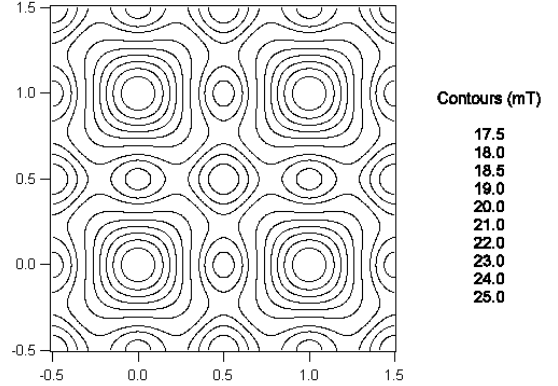}}
  \caption[~]{Contour plot of the magnetic field in 
the mixed state of SRO as reconstructed from the data represented 
in Figure~\ref{fig:fll}, using the signs of the Fourier components as discussed 
in the text.} \label{fig:reconstructed}
\end{figure}
\begin{table}[t]
\caption[~]{Calculated and experimental flux lattice Fourier components
and intensities for $B$=20mT, $B_{c2}$=58mT .}\label{tab:tb1}
  \begin{center}
    \begin{tabular}{cccc}
      $h,k$ of diffraction peak: & \{1,1\}  &  \{2,0\}  &  \{2,1\} \\
      \hline
      $q_{hk}/q_{10}$ & 1.414 & 2.0   & 2.236\\
      $F_{hk}/F_{10}$ (London $\lambda=152nm):$ & 0.53 & 0.27 & 0.22\\
      $F_{hk}/F_{10}$ (Abrikosov): & 0.2079 & -0.00898 & 0.00187 \\
      $F_{hk}/F_{10}$ (Agterberg $\nu =0.0$): & 0.5657  & -0.1051 & 0.0353 \\
      $F_{hk}/F_{10}$ (Agterberg $\nu =0.2$): & 0.7711 & -0.0667 & 0.0503 \\
      $F_{hk}/F_{10}$ (Heeb and Agterberg & 0.484 & -0.019 & 0.046 \\
      $\nu =0.05$, $\kappa = 1.6$): & & & \\
      $I_{hk}/I_{10}$ (London): & 0.199 & 0.0365 & 0.0216 \\
      $I_{hk}/I_{10}$ (Abrikosov): & 0.0306 & 0.00004 & 0.000002 \\
      $I_{hk}/I_{10}$ (GL (Brandt)): & 0.0783 & 0.00091 & 0.000298 \\
      $I_{hk}/I_{10}$ (Agterberg $\nu =0.0$): & 0.2263 & 0.00552 & 0.00056 \\
      $I_{hk}/I_{10}$ (Heeb and Agterberg & 0.166 & 0.00095 &0.00018 \\
      $\nu =0.05$, $\kappa=1.6$): & & & \\ 
      $I_{hk}/I_{10}$ & 0.197(2) & 0.011(3) & 0.007(1)\\
      (Expt. at 20 mT 100mK): & & & \\
    \end{tabular}
  \end{center}
\end{table}
\begin{table}[htbp]
\caption[~]{Experimental FLL Fourier components at 100mK.  Errors represent 
statistical errors; FLL disorder or static Debye Waller 
factors may reduce $F_{hk}$ for high ${\mathbf{q}}$.}\label{tab:tb2}
  \begin{center}
    \begin{tabular}{ccc}
      Magnetic Field (mT) & $F_{10}$\,(mT) & $F_{11}$\,(mT) \\
      \hline
      100 & 1.14(3) & 0.66(3) \\
      150 & 1.02(9) & 0.54(1) \\
      200 & 0.88(1) & 0.47(1) \\
      250 & 0.71(1) & 0.36(1) \\
      300 & 0.58(1) & 0.29(1) \\
      350 & 0.45(1) & 0.24(1) \\
      400 & 0.30(2) & 0.14(1) \\
    \end{tabular}
  \end{center}
\end{table}

\end{document}